\documentclass{article}

    \usepackage[final,nonatbib]{neurips_2019}

\usepackage[utf8]{inputenc} 
\usepackage[T1]{fontenc}    
\usepackage{hyperref}       
\usepackage{url}            
\usepackage{booktabs}       
\usepackage{amsfonts}       
\usepackage{nicefrac}       
\usepackage{microtype}      
\usepackage{color}          
\usepackage{graphicx}
\usepackage{multirow}

\newcommand{\mlmodel}{R3Net}  
\newcommand{\recvrate}{\mathcal{R}}
\newcommand{\delay}{\mathcal{D}}
\newcommand{\lossrate}{\mathcal{L}}

\title{Reinforcement learning for bandwidth estimation and congestion control in real-time communications}

\author{%
Joyce Fang,
Martin Ellis,
Bin Li,
Siyao Liu,
Yasaman Hosseinkashi,
Michael Revow,\\
{\bf 
Albert Sadovnikov,
Ziyuan Liu,
Peng Cheng,
Sachin Ashok,
}\\
{\bf
David Zhao,
Ross Cutler,
Yan Lu,
Johannes Gehrke
}
\\\\
Microsoft
}

\begin{document}

\maketitle

\begin{abstract}
  Bandwidth estimation and congestion control for real-time communications (i.e., 
audio and video conferencing) remains a difficult problem, despite many years of research.
Achieving high quality of experience (QoE) for end users requires continual updates
due to changing network architectures and technologies.
In this paper, we apply reinforcement learning for the first time to the problem of real-time communications (RTC), where we seek to optimize user-perceived quality.
We present initial proof-of-concept results, where we learn an agent to control sending rate in an RTC system, evaluating using both network simulation and real Internet video calls.
We discuss the challenges we observed, particularly in designing realistic reward functions that reflect QoE, and in bridging the gap between the training environment and real-world networks.

\end{abstract}

\section{Introduction}
\label{s:introduction}

Congestion control and bandwidth estimation are fundamental problems in networking research.
They are concerned with how much data can be sent across a network path at a given time, and how and when endpoints should send packets to avoid causing network congestion and the associated packet delay and loss~\cite{winstein2013remycc}.
There have been various applications of reinforcement learning (RL) to these problems, such as for congestion control in TCP~\cite{jay2019pccrl, ruffy2018iroko} and QUIC~\cite{sivakumar2019mvfstrl}, and one-way adaptive bitrate video streaming~\cite{mao2017neural, huang2018qarc, yan2019continual, bhattacharyya2019qflow}.
In this paper, we apply RL to congestion control and bandwidth estimation in real-time communications (RTC), for the first time.
We begin by reviewing prior applications of RL to video streaming (since it is most closely related to RTC) before outlining the differences in our approach.

\textbf{RL for Video Streaming:}
RL has been successfully applied to the control of adaptive bitrate video streaming systems (i.e., YouTube-like one-way video streaming), with results showing that RL gives large improvements over existing approaches under certain conditions~\cite{mao2017neural, huang2018qarc, bhattacharyya2019qflow}.
In these systems, clients typically select videos to download from a fixed set of available quality levels, which correspond to discrete actions for RL agents~\cite{mao2017neural}.
RL reward functions can be designed based on QoE metrics, combining measurements such as bitrate, delay, and video quality with configurable weights.
The most similar domain to RTC is \emph{real-time} video streaming~\cite{huang2018qarc}, which can be thought of as ``one-directional'' high latency RTC.
This has some of the real-time constraints of RTC, although since it is non-interactive, the requirements are less strict.

\textbf{RL for RTC:}
Beyond video streaming, an equally important but harder problem is  real-time communications (RTC).
In WebRTC~\cite{narayanan2018webrtc1.0}, several congestion control approaches have been proposed~\cite{carlucci2016gcc, zhu2013nada, johansson2014scream};
other work has focused on optimizing RTC performance beyond WebRTC~\cite{winstein2013stochastic, fouladi2018salsify}. 
To the best of our knowledge, RL has not been applied to this space.

RTC is different from video streaming for the following reasons.
First, RTC requires minimal latency and cannot pre-fetch content, so large receiver-side buffers (common in video streaming) cannot be used; the system needs to react faster to bandwidth changes with less margin for error.
This constraint also means that packet losses have a bigger impact, since there is less time available to retransmit lost packets. 
Second, since RTC involves end users uploading their audio/video streams, quality is likely be limited by their uplink capacity, which is often more constrained than their downlink capacity.
Third, since the RL model needs to run in a real-time environment, the inference time needs to be orders of magnitude faster than the streaming case, further constraining the complexity of the RL model.
Fourth, since RTC systems do not work with the pre-encoded quality levels that are typical in video streaming systems, the action space in RTC is typically larger or continuous. 

Due to the constraints above, we cannot apply previous RL formulations designed for one-directional delivery with high buffer latency.
To address these challenges, we propose \mlmodel{}, an RL-based Recurrent Network for RTC, allowing rapid adjustment to complex and dynamic network conditions. 

\textbf{Paper Outline:}
We outline our initial training environment, simulator, and model
in \S \ref{s:rl-approach}.
We evaluate the model (using simulation and video calls on real networks) in \S \ref{s:evaluation}, 
and discuss open issues in \S \ref{s:discussion}.

\section{\mlmodel{}: An Initial Approach}
\label{s:rl-approach}

In RL the formulation of the problem in terms of states, actions, and reward is crucial.
In RTC, the ultimate reward is to deliver excellent QoE to end users, although the actions that can be taken to achieve this can vary widely.
In our present work, we focus on a subset of the problem (bandwidth estimation and congestion control), but we note that there are numerous other sub-problems in RTC can naturally be posed as RL problems (e.g., jitter-buffer control, packet loss resiliency, video encoding, etc). 
Eventually, an RL agent might control all actions in an RTC system, continuously improving QoE in an online manner.

We take a receiver-side approach to bandwidth estimation in RTC calls, using incoming RTP~\cite{rfc3550} packets to estimate available bandwidth on the path between sender and receiver.
We then signal this estimate back to the sender via RTCP, allowing sender-side logic to control sending rate.
Our existing reference system uses an Unscented Kalman Filter (UKF) with a rule-based controller to estimate and control bandwidth.
Our initial approach to applying RL to RTC is to use the observations of the incoming packet timeseries as input to the neural network, training a model to estimate the available bandwidth that will replace the UKF method. 
We will compare the methods in \S \ref{s:evaluation}.

\subsection{Simulator}
\label{ss:simulation}

In RL training, it is common to use a simulation environment to speed up training, allowing agents to learn from vast numbers of observations before they are deployed into their target environment. 
For RTC, this means that a realistic simulation of Internet and application performance is required. Training may also be done online, with observations being collected from a full-scale RTC system, even learning from production calls.
In the latter case, the collection of training data and real-time continuous updating of the model needs to be considered.

We can train an RL model either in the real RTC process or in a simulator.
As an initial approach, we use a simulator that can mimic the RTC process, but runs 1000x faster than real-time, to speed up training.
Our simulator consists of the caller and callee RTC endpoints connected by a simulated network link; we can also simulate cross traffic (e.g., from TCP senders).
The network simulator uses trace-replay-based simulation to control the parameters of the bottleneck link (including capacity, delay, and packet loss) in a discrete event simulation.

\subsection{RL formulation}
\label{ss:rl-formuation}
To estimate bandwidth, the receiver can use the incoming RTP packets and the measured round-trip time (RTT).
In our formulation, we use the aggregated RTP and RTT information in a fixed time-window of 50 ms as the environment state, and estimated bandwidth as the agent's action.
The environment updates the next state and reward based on the input action.

\textbf{State and Action:}
The state is a 4-dimension vector consisting of receive rate (kb/s), average packet interval (ms), packet loss rate (\%), and average RTT (ms).
We further scale the state to produce inputs with the same order of magnitude to the neural network.
We use sigmoid activation as the last layer of the network, yielding outputs in the $(0, 1)$ range.
We then map the output to $(0, 8)$ Mb/s as the bandwidth estimate, corresponding to an appropriate range for our RTC application.

\textbf{Reward Design:}
We define reward per 50ms time step as 
$0.6 \ln(4 \recvrate{} + 1) - \delay{} - 10 \lossrate{}$,
where 
$\recvrate{}$ is receive rate in that time step, in Mb/s,
$\delay{}$ is the average RTT in that time step, in seconds, and  
$\lossrate{}$ is packet loss rate.
This means that receiving more packets is rewarded (since this should lead to higher QoE), but delay and packet loss are penalized (since these degrade QoE).

\subsection{Model and Training}
\label{ss:model-training}

\begin{figure}[t]
    \centering
    \includegraphics[trim=10 20 20 20,clip,width=\textwidth]{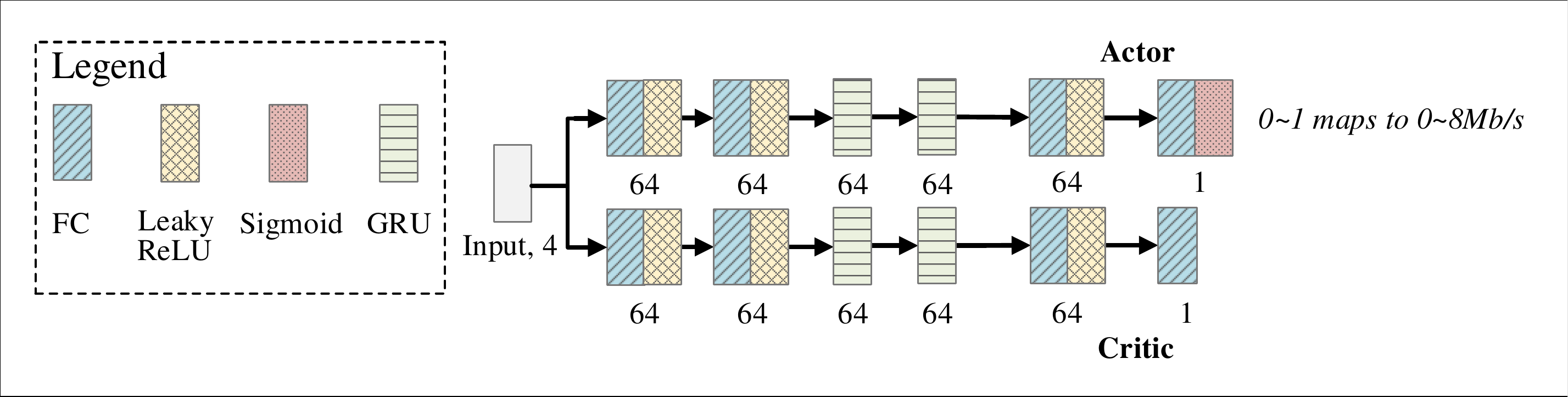}
    \caption{\mlmodel{} structure (numbers indicate output features per layer, and hidden size for GRUs)}
    \label{f:network-structure}
\end{figure}

The input of the neural network is a time series, representing the state of the path between sender and receiver over time.
The history information has impact on the estimated bandwidth (e.g., increasing RTT may mean previous bandwidth estimates were too large).
Thus, we use a recurrent neural network with Gated Recurrent Units (GRUs)~\cite{Cho2014GRU} to estimate bandwidth, as shown in Figure \ref{f:network-structure}.
For the leaky ReLU layer, we use the negative slope of 0.01.
For the rest of the paper, we refer to this neural network as \mlmodel{} (RL-based Recurrent Network for RTC).

We train \mlmodel{} using an actor-critic framework, where the actor and critic share the first few layers.
The model is updated using Proximal Policy Optimization (PPO)~\cite{Schulman201707PPO} and the Adam optimizer with a learning rate of $3\times10^{-5}$, implemented using PyTorch, based on DeepRL \cite{deeprl}.
We used around 10,000 network traces for simulation in training, and tested on 1150 different network traces.

\section{Evaluation}
\label{s:evaluation}

\subsection{Evaluation Through Simulation}
\label{ss:evaluation-simulation}

\begin{figure}[t]
    \centering
    \includegraphics[trim=0 20 0 5,clip,width=\textwidth]{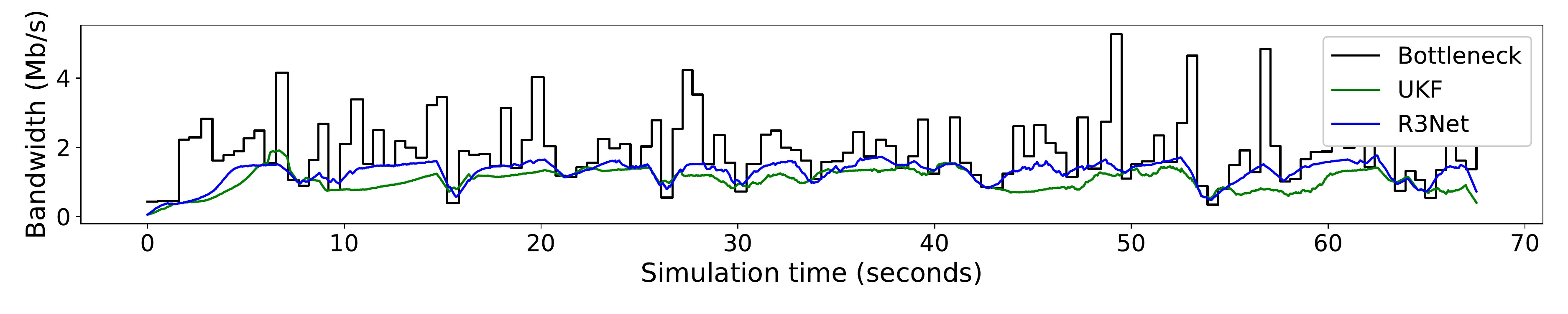}
    \caption{Example result from one simulation test run}
    \label{f:example-result}
\end{figure}

We first evaluate \mlmodel{} in simulation, comparing \mlmodel{} with UKF based on our set of 1150 test traces.
Ideally, our evaluation criteria would be based on user-perceived quality (e.g., MOS~\cite{itut-p.800.1, itut-p.800}), but since our simulation environment uses synthetic audio and video packets, we use purely network-based metrics including observed RTT, packet loss rate, and \emph{bandwidth utilization}, the percentage of bandwidth used relative to the (simulated) limit.
Detailed results are shown in Table \ref{t:simulation-results}; we see that \mlmodel{} has $\sim$5\% higher bandwidth utilization than UKF (see Figure \ref{f:example-result} for an example), with similar RTTs and less packet loss.
These initial simulation results are promising, with \mlmodel{} showing higher reward and better overall performance than UKF; in the next section, we evaluate model performance in real network conditions.

\begin{table}[t]
    \caption{Evaluation results from simulation}
    \label{t:simulation-results}
    \centering
    \begin{tabular}{|c||c|c|c|c|c|c|}
    \hline
                & \multirow{2}{*}{Bandwidth utilization}  & \multicolumn{3}{c|}{RTT (ms)}      & \multirow{2}{*}{Packet loss rate} & \multirow{2}{*}{Reward mean}\\
                &             & avg. &  p50 &  p95 &  &\\
    \hline
    UKF         & 73.5\%      &  128 &  102 &  288 &  0.38\% & 0.56\\
    \mlmodel{}  & 77.8\%      &  122 &  102 &  268 &  0.19\% & 0.60\\
    \hline
    \end{tabular}
\end{table}

\subsection{Evaluation Using Real Networks}
\label{ss:evaluation-real}

We now describe our very preliminary evaluation of \mlmodel{} performance in RTC calls on real networks.
%
First, we deployed the \mlmodel{} into the ONNX format~\cite{onnx}, and use ONNX Runtime~\cite{onnxruntime} for inference in our RTC system.
Currently the inference time of \mlmodel{} takes approximately 500 $\mu$s and the model is called every 50 ms.
The inference time of \mlmodel{} is about 20 times more expensive than UKF and rule-based approaches, but is still within the runtime requirement.
%
To compare the performance of UKF and \mlmodel{}, we ran two-way audio/video calls using a scriptable RTC client in two different scenarios, \emph{WiFi}, and \emph{3G}, where the first machine is connected via residential WiFi and USB-tethered 3G connections, respectively.
In all scenarios, the second machine is connected to our office network, and has the RTC application record the incoming video stream, allowing us to compute objective quality scores using VMAF~\cite{vmaf}.
This allows us to objectively evaluate the performance of UKF and \mlmodel{}, independently of the training environment.

%
We ran 200 test calls for \emph{3G} and 200 for \emph{WiFi}, each lasting 30 seconds, alternating between running UKF and \mlmodel{}.
Table \ref{t:real-network-results} shows the RTT, packet loss rate, and video quality scores.
We see that although the RTTs are fairly similar between \mlmodel{} and UKF for each network type, packet loss rates are higher on 3G networks when using \mlmodel{}.
There are corresponding degradations in both VMAF (indicating poorer image quality) and video frame drop rate (indicating choppy video).
This suggests that the simulation environment does not sufficiently represent the real network environment.
We observe that \mlmodel{} takes relatively noisy actions compared to UKF, which might lead to high packet loss and choppy video in real networks.

\begin{table}[t]
    \caption{Evaluation results from 3G and WiFi}
    \label{t:real-network-results}
    \centering
    \begin{tabular}{|c||c|c|c|c|c|c|c|}
    \hline
                & \multirow{2}{*}{Network type} & \multicolumn{3}{c|}{RTT (ms)}  & \multirow{2}{*}{Packet loss rate} & \multirow{2}{*}{VMAF} & \multirow{2}{*}{Frame drop rate} \\
                &  & avg. & p50 & p95 & & &  \\
    \hline
    \multirow{2}{*}{UKF}          & 3G   &  58 &  56 & 86 &  2.22\%  &  81.8  & 6.5\% \\
                                  & WiFi &  16 &  13 &  37 &  0.05\%  &   94.1  & 2.5\% \\
    \hline
    \multirow{2}{*}{\mlmodel{}}   & 3G   & 58 &  55 & 99 & 3.11\%  &   78.6  & 11.2\% \\
                                  & WiFi &  16 &  13 &  38 &  0.01\%  &   93.4  & 1.8\%  \\
    \hline
    \end{tabular}
\end{table}

\section{Discussion and Open Questions}
\label{s:discussion}

In this paper, we propose a new formulation of RL for bandwidth estimation and congestion control in real-time audio/video communication, and show \mlmodel{} provides reasonable adjustment to dynamic network conditions in simulation and real networks using WiFi connections.
Although the evaluation results in 3G networks suggest further improvement of the model is needed, we hope our end-to-end training and deployment of RL to RTC stimulates further work in this direction.
We identify two key areas for improvement:
1) \emph{How to close the gap between training and the real world through realistic network simulation?}
2) \emph{How to formulate reward functions that directly optimize QoE?}

From our preliminary results, we are encouraged that \mlmodel{} can deliver a reasonable experience over WiFi (though not yet matching UKF).
However, since \mlmodel{} suffers high packet loss (and poor video quality) in our 3G tests, it is clear our training environment is not sufficiently representative.
It is common to start RL training in a simulator (i.e., a gym environment), but developing a sufficiently realistic environment to represent the real world is challenging.
In future work, we plan to improve the simulation using data driven methods (i.e., using a generative model to produce realistic network traces), and to build a distributed testbed to enable large scale training using real networks.

In our training, we used a simple intuitive reward, and see that \mlmodel{} performs more aggressively (i.e., uses more bandwidth) than UKF in simulation-based evaluation.
The higher packet loss rate in real network evaluation shows that we need to redesign this reward function.
We believe that using objective functions for audio/video quality measured throughout the call as multi-step rewards in training may yield better real-world performance; the design of a reward function that leads to high QoE is a key challenge for future work.

\subsubsection*{Acknowledgments}

We would like to thank Nelson Pinto for help in developing the simulation environment, 
Sergiy Matusevych for help with ONNX Runtime integration, 
James Winters and Yang Sun for their work building the evaluation environment, and 
Sasikanth Bendapudi for help with video quality assessment.

\bibliographystyle{abbrv}
\begin{small}
\bibliography{reference}
\end{small}

\end{document}